\documentclass[twocolumn,prl]{revtex4}

\usepackage{dcolumn}
\usepackage{amsmath}

\usepackage{graphicx}

\begin{document}

% Macros
\newcommand{\defn}{\textit}
\newcommand{\eref}[1]{(\ref{#1})}
\newcommand{\etal}{{\it{}et~al.}}

% Style parameters
\newlength{\figurewidth}
\setlength{\figurewidth}{0.95\columnwidth}
\setlength{\parskip}{0pt}
\setlength{\tabcolsep}{3.5pt}

\title{Shape and efficiency in spatial distribution networks}
\author{Michael T. Gastner}
\affiliation{Department of Physics, University of
Michigan, Ann Arbor, MI 48109}
\author{M. E. J. Newman}
\affiliation{Department of Physics, University of
Michigan, Ann Arbor, MI 48109}
\begin{abstract}
We study spatial networks that are designed to distribute or collect a
commodity, such as gas pipelines or train tracks.  We focus on the cost of
a network, as represented by the total length of all its edges, and its
efficiency in terms of the directness of routes from point to point.  Using
data for several real-world examples, we find that distribution networks
appear remarkably close to optimal where both these properties are
concerned.  We propose two models of network growth that offer explanations
of how this situation might arise.
\end{abstract}
\maketitle

A network is a set of points or \defn{vertices} joined together in pairs by
lines or \defn{edges}.  Networks provide a useful framework for the
representation and modeling of many physical, biological, and social
systems, and have received a substantial amount of attention in the recent
physics literature~\cite{AB02,DM02,Newman03d}.  In this paper we study
networks in which the vertices occupy particular positions in geometric
space.  Not all networks have this property---web pages on the world wide
web, for example, do not live in any particular geometric space---but many
others do.  Examples include transportation networks, communication
networks, and power grids. Recently several studies have appeared in the
physics literature that address the ways in which geography influences
networks~\cite{YJB02,GK04,GMTA04,CS04,KH04,GN05}.

In this paper we study the spatial layout of man-made distribution or
collection networks, such as oil and gas pipelines, sewage systems, and
train or air routes.  The vertices in these networks represent, for
instance, households, businesses, or train stations and the edges represent
pipes or tracks.  In most cases the network also has a ``root node'', a
vertex that acts as a source or sink of the commodity distributed---a
sewage treatment plant, for example, or a central train station.

Geography clearly affects the efficiency of these networks.  A ``good''
distribution network as we will consider it in this paper has two
definitive properties.  First, the network should be efficient in the sense
that the paths from each vertex to the root vertex are relatively short.
That is, the sum of the lengths of the edges along the shortest path
through the network should be not much longer than the ``crow flies''
distance between the same two vertices: if a subway track runs all around
the city before getting you to the central train station, the train is
probably not of much use to you.  Second, the sum of the lengths of all
edges in the network should be low so that the network is economical to
build and maintain.  In this paper we argue that these two criteria are
often at odds with one another, but that even so, real networks manage to
find solutions to the distribution problem that come remarkably close to
being optimal in both senses.  We suggest possible explanations for this
observation in the form of two growth models for geographic networks that
generate networks of comparable efficiency to our real-world examples.

We begin our study by looking at the properties of some real-world
distribution networks.  We consider four examples as follows.

Our first network is the sewer system for the City of Bellingham,
Washington.  From GIS data for the city we extracted the shapes and
positions of the parcels of land (roughly households) into which the city
is divided and the lines along which sewers run.  We constructed a network
by assigning one vertex to each parcel whose centroid was less than 100
meters from a sewer.  The vertex was placed on the sewer at the point
closest to the corresponding centroid and adjacent vertices along the
sewers were connected by edges.  The city's sewage treatment plant was used
as the root vertex, for a total of $23\,922$ vertices including the root.

Our next two examples are networks of natural gas pipelines, the first in
Western Australia (WA) and the second in the southeastern part of the US
state of Illinois (IL)~\footnote{South of $41.00^\circ$N and east of
$89.85^\circ$W.  We consider only the largest component within this
region.}.  We assigned one vertex to each city, town, or power station
within $10$km (WA) or 10,000 feet (IL) of a pipeline.  The vertex was
placed on the pipeline at the point closest to each such place, and
adjacent vertices joined by edges.  The root for WA was chosen to be the
shore point of the pipeline leading to the Barrow Island oil fields and for
IL to be the confluence of two major trunk lines near the town of
Hammond,~IL.  The resulting networks have $226$ (WA) and $490$ (IL)
vertices including the roots.

\begin{figure*}
\begin{center}
\resizebox{16cm}{!}{\includegraphics{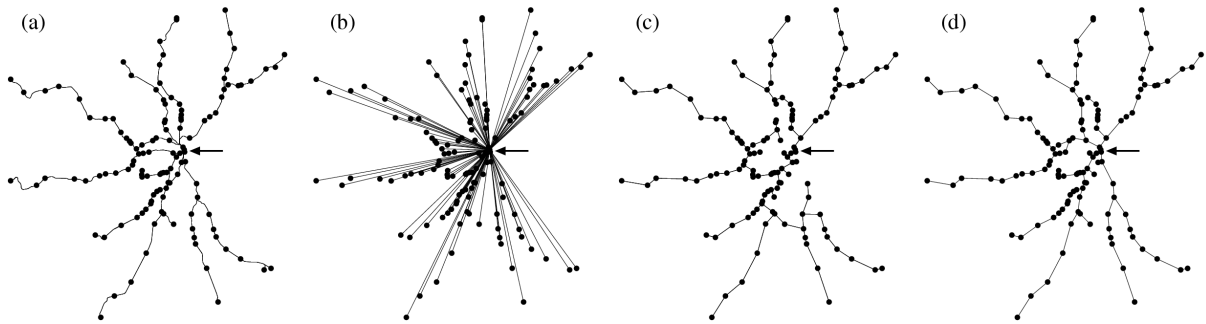}}
\end{center}
\caption{(a) Commuter rail network in the Boston area.  The arrow marks the
assumed root of the network.  (b)~Star graph. (c)~Minimum spanning tree.
(d)~The model of Eq.~\eref{betamodel} applied to the same set of stations.}
\label{MBTA}
\end{figure*}

For our last example we take the commuter rail system operated by the
Massachusetts Bay Transportation Authority in the city of Boston, MA
(Fig.~\ref{MBTA}a).  In this network, the 125 stations form the vertices
and the tracks form the edges.  In principle, there are two components to
this network, one connected to Boston's North Station and the other to
South Station, with no connection between the two.  Since these two
stations are only about one mile apart, however, we have, to simplify
calculations, added an extra edge between the North and South Stations,
joining the two halves of the network into a single component.  The root
node was placed halfway between the two stations for a total of 126
vertices in all.

We wish to quantify the efficiency of these networks in terms of path
lengths and combined edge length, as described above.  To do this, we
compare our measurements of the networks to two theoretical models that are
each optimal by one of these two criteria.  If one is interested solely in
short, efficient paths to the root vertex then the optimal network is the
``star graph,'' in which every vertex is connected directly to the root by
a single straight edge (see Fig.~\ref{MBTA}b).  Conversely, if one is
interested solely in minimizing total edge length, then the optimal network
is the minimum spanning tree (MST) (see Fig.~\ref{MBTA}c).  (Given a set of
$n$ vertices at specified points on a flat plane, the MST is the set of
$n-1$ edges joining them such that all vertices belong to a single
component and the sum of the lengths of the edges is minimized~\footnote{If
we are not restricted to the specified vertex set but are allow to add
vertices freely, then the optimal solution is the Steiner tree; in practice
we find that there is very little difference between results for minimum
spanning and Steiner trees in the present context.}.)

To make the comparison with the star graph, we consider the distance from
each non-root vertex to the root first along the edges of the network and
second along a simple Euclidean straight line, and calculate the mean ratio
of these two distances over all such vertices.  Following
Ref.~\cite{Black03}, we refer to this quantity as the network's \defn{route
factor}, and denote it~$q$:
\begin{equation}
q = \frac{1}{n}\sum_{i=1}^n\frac{l_{i0}}{d_{i0}},
\label{routefac}
\end{equation}
where $l_{i0}$ is the distance along the edges of the network from vertex
$i$ to the root (which has label~0), and $d_{i0}$ is the direct Euclidean
distance.  If there is more than one path through the network to the root,
we take the shortest one.  Thus, for example, $q=2$ would imply that on
average the shortest path from a vertex to the root through the network is
twice as long as a direct straight-line connection.  The smallest possible
value of the route factor is~1, which is achieved by the star graph.

\begin{table}
\begin{tabular}{l|r|rr|rrr}
\multicolumn{2}{c|}{ } & \multicolumn{2}{c|}{route factor}
                       & \multicolumn{3}{c}{edge length (km)} \\
network                      & $n$       & actual & MST    & actual & MST & star \\
\hline
sewer system & $23\,922$ & $1.59$ & $2.93$ &     $498$ &    $421$ & $102\,998$ \\
gas (WA)     &     $226$ & $1.13$ & $1.82$ &  $5\,578$ & $4\,374$ & $245\,034$ \\
gas (IL)     &     $490$ & $1.48$ & $2.42$ &  $6\,547$ & $4\,009$ &  $59\,595$ \\
rail         &     $126$ & $1.14$ & $1.61$ &     $559$ &    $499$ &   $3\,272$
\end{tabular}
\caption{Number of vertices~$n$, route factor~$q$, and total edge length
for each of the networks described in the text, along with the equivalent
results for the star graphs and minimum spanning trees on the same
vertices.  (Note that the route factor for the star graph is always~1 and
so has been omitted from the table.)}
\label{summary}
\end{table}

The route factors for our four networks are shown in Table~\ref{summary}.
As we can see, the networks are remarkably efficient in this sense, with
route factors quite close to~1.  Values range from $q=1.13$ for the Western
Australian gas pipelines to $q=1.59$ for the sewer system.

We also show in Table~\ref{summary} the total edge lengths for each of our
networks, along with the edge lengths for the MST on the same set of
vertices and, as the table shows, we again find that our real-world
networks are competitive with the optimal model, the combined edge lengths
of the real networks ranging from $1.12$ to $1.63$ times those of the
corresponding MSTs.

But now consider the remaining two columns in the table, which give the
route factors for the MSTs and the total edge lengths for the star graphs.
As the table shows, these figures are for all networks much poorer than the
optimal case and, more importantly, much poorer than the real-world
networks too.  Thus, although the MST is optimal in terms of total edge
length it is very poor in terms of route factor and the reverse is true for
the star graph.  Neither of these model networks would be a good general
solution to the problem of building an efficient and economical
distribution network.  Real-world networks, on the other hand, appear to
find a remarkably good compromise between the two extremes, possessing
simultaneously the benefits of both the star graph and the minimum spanning
tree, without any of the flaws.  In the remainder of the paper we consider
mechanisms by which this might occur.

The networks we are dealing with are not, by and large, designed from the
outset for global optimality (or near-optimality) of either their total
edge length or their route factors.  Instead, they form by growing outward
from the root, as the population they serve swells and infrastructure is
extended and improved.  To explore the possibilities of this process we
consider a situation in which the positions of vertices (houses, towns,
etc.)\ are given and we are to build a network connecting them.  For
simplicity we will initially assume that the vertices are randomly
distributed in two-dimensional space with unit mean density, with one
vertex designated as the root of the network.  A cluster connected to the
root is built up by repeatedly adding an edge that joins one unconnected
vertex~$i$ to another~$j$ that is part of the cluster.  The question is how
these edges are to be chosen.  Our proposal is to use a simple greedy
optimization criterion.

We specify a weight for each edge~$(i,j)$ thus:
\begin{equation}
w_{ij} = d_{ij} + \alpha\frac{d_{ij}+l_{j0}}{d_{i0}},
\label{alphamodel}
\end{equation}
where $\alpha$ is a non-negative independent parameter.  As before,
$d_{ij}$ is the direct Euclidean distance between vertices $i$ and~$j$ and
$l_{ij}$ the distance along the shortest path in the network.  The first
term in~\eref{alphamodel} is the length of the prospective edge, which
represents the cost of building the corresponding pipe or track, and the
second term is the contribution to the route factor from vertex~$i$.  At
every step we now add to the network the edge with the global minimum value
of $w_{ij}$.  The single parameter $\alpha$ controls the extent to which
our choice of edge depends on the route factor.  For $\alpha=0$ we always
add the vertex that is closest to the connected cluster.  This limit
produces a graph akin to a grown version of the minimum spanning tree, and
we find it to give very poor route factors.  As $\alpha$ is increased from
zero, however, the model becomes more and more biased in favor of making
connections that give good values for the route factor.

\begin{figure}
\begin{center}
\resizebox{\figurewidth}{!}{\includegraphics{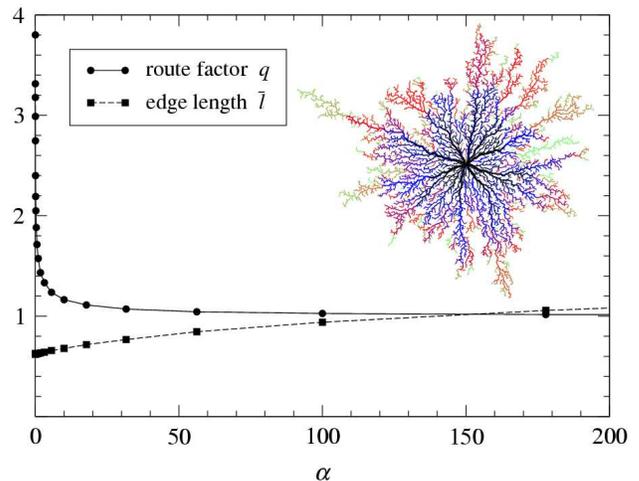}}
\end{center}
\caption{Simulation results for the route factor $q$ and average edge
length $\bar{l}$ as a function of~$\alpha$ for our first model with
$n=10\,000$ vertices.  Inset: an example model network with $\alpha=12.0$.
Colors indicate the order in which edges were added to the network.}
\label{alphafig}
\end{figure}

Figure~\ref{alphafig} shows results from simulations of this model.  We
plot the route factor~$q$ of the entire network and the average length of
an edge $\bar{l}$ against~$\alpha$.  As $\alpha$ is increased the route
factor does indeed go down in this model, just as we expect.  What is
interesting however is that $q$ initially decreases very sharply
with~$\alpha$, while at the same time~$\bar{l}$, which is a measure of the
cost of building the network, increases only slowly.  Thus it appears to be
possible to grow networks that cost only a little more than the optimal
($\alpha=0$) network, but which have far less circuitous routes.  This
finding fits well with our observations of real distribution networks.

The inset to Fig.~\ref{alphafig} shows an example network grown using this
model.  The network has a dendritic appearance, with relatively straight
trunk lines and short branches, and bears a qualitative resemblance to
diffusion-limited aggregation clusters~\cite{WS81} or dielectric breakdown
patterns~\cite{NPW84}, which have also been used as models of urban
growth~\cite{BLF89} although they are based on entirely different
mechanisms.

In some respects, however, this model is quite unrealistic.  In particular,
many vertices are never joined to the network, even ones lying quite close
to the root, because to do so would simply be too costly in terms of the
route factor.  (This is the reason for the dendritic shape.)  This is not
the way the real world works: one doesn't decide not to provide sewer
service to some parts of a city just because there's no convenient straight
line for the sewer to take.  Instead, connections seem to be made to those
vertices that can be connected to the root by a reasonably short path,
regardless of whether that path is straight.  In the case of trains, for
instance, people will use a train service---and thereby justify its
construction---if their train journey is short in absolute terms, and are
less likely to take a longer journey even if the longer one is along a
straighter line.  As we now show, we can, by incorporating these
considerations, produce a more realistic model that still generates highly
efficient networks.

\begin{figure}
\begin{center}
\resizebox{\figurewidth}{!}{\includegraphics{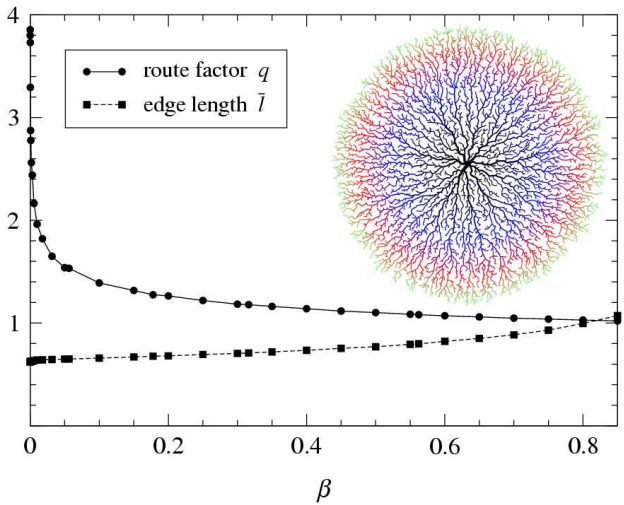}}
\end{center}
\caption{Route factor $q$ and average edge length $\bar{l}$ as a function
of~$\beta$ for our second model ($n=10\,000$).  Inset: an example model
network with $\beta=0.4$.}
\label{betafig}
\end{figure}

Let us modify Eq.~\eref{alphamodel} to give preference to short paths
regardless of shape.  To do this, we write the weight of a new edge~$(i,j)$
as simply
\begin{equation}
w_{ij}' = d_{ij} + \beta l_{j0}.
\label{betamodel}
\end{equation}
(A model with a similar weight function was previously studied by
Fabrikant~\etal~\cite{FKP02}, but gives quite different results from ours
because vertices were added to the network one by one, rather than being
specified from the outset as in our case.) Note that there is now no
explicit term that guarantees low route factors.  Nonetheless, the model
self-organizes to a state whose route factor is small.
Figure~\ref{betafig} shows results from our simulations of this second
model.  As the plot shows, the results are qualitatively quite similar to
our first model: the high value of~$q$ seen for $\beta=0$ drops off quickly
as $\beta$ is increased, while the mean edge length increases only slowly.
Thus we can again choose a value for~$\beta$ that gives behavior comparable
with our real-world networks, having simultaneously low route factor and
low total cost of building the network.  Values of $q$ in the range $1.1$
to $1.6$ observed in the real-world networks are easily achieved.

When we look at the shape of the network itself however (see figure inset),
we get quite a different story.  This model produces a symmetric network
that fills space out to some approximately constant radius from the root,
not unlike the clusters produced by the well-known Eden growth
model~\cite{Eden61}.  The second term in Eq.~\eref{betamodel} makes it
economically disadvantageous to build connections to outlying areas before
closer areas have been connected.  Thus all vertices within a given
distance of the root are served by the network, without gaps, which is a
more realistic situation than the dendritic network of Fig.~\ref{alphafig}.

And this in fact may be the secret of how low route factors are achieved in
reality.  Our second model---unlike our first---does not explicitly aim to
optimize the route factor.  But it does a creditable job nonetheless,
precisely because it fills space radially.  The main trunk lines in the
network are forced to be approximately straight simply because the space to
either side of them has already been filled and there's nowhere else to go
but outwards.

Readers familiar with urban geography may argue that real networks, and the
towns they serve, \emph{are} dendritic in form.  And this is true, but it
is primarily a consequence of other factors, such as ribbon development
along highways.  In other words, the initial distribution of vertices in
real networks is usually non-uniform, unlike our model.  It is interesting
to see therefore what happens if we apply our model to a realistic scatter
of points, and in Fig.~\ref{MBTA}d we have done this for the stations of
the Boston rail system.  The figure shows the network generated by our
second model for $\beta=0.4$ given the real-world positions of the
stations.  The result is, with only a couple of exceptions, identical to
the true rail network, with a comparable route factor of $1.11$ and total
edge length $511$km.

To summarize, we have in this paper studied spatial distribution or
collection networks such as pipelines and sewers, focusing particularly on
their cost in terms of total edge length and their efficiency in terms of
the network distance between vertices, as measured by the so-called route
factor.  While these two quantities are, to some extent, at odds with one
another, the first being decreased only at the expense of an increase in
the second, our empirical observations indicate that real-world networks
find good compromise solutions giving nearly optimal values of both.  We
have presented two models of spatial networks based on greedy optimization
strategies that reproduce this behavior well, showing how networks
possessing simultaneously good route factors and low total edge length can
be generated by plausible growth mechanisms.

The results presented represent only a fraction of the possibilities in
this area.  Numerous other networks fall into the class studied here,
including various utility, transportation, or shipping networks, as well as
some biological networks, such as the circulatory system, fungal mycels,
and others, and we hope that researchers will feel encouraged to
investigate these interesting systems.

The authors thank Jonathan Goodwin and Sean Doherty for the pipeline
network data and the staff of the University of Michigan's Numeric and
Spatial Data Services for their help.  This work was funded in part by the
National Science Foundation under grant number DMS--0234188 and by the
James S. McDonnell Foundation.

\end{document}